\documentstyle[11pt,amsfonts]{article}
\renewcommand{\theequation}{\arabic{equation}}
\newcommand{\be}{\begin{equation}}
\newcommand{\ee}{\end{equation}}
\newcommand{\bea}{\begin{array}}
\newcommand{\ea}{\end{array}}
\newcommand{\beqa}{\begin{eqnarray}}
\newcommand{\eeqa}{\end{eqnarray}}
\newcommand{\bean}{\begin{eqnarray*}}
\newcommand{\eean}{\end{eqnarray*}}

\def\up#1{\leavevmode \raise.16ex\hbox{#1}}

\newcommand{\gapproxeq}{\lower
 .7ex\hbox{$\;\stackrel{\textstyle >}{\sim}\;$}}
\newcommand{\lapproxeq}{\lower .7ex\hbox{$\;\stackrel
{\textstyle <}{\sim}\;$}}

\renewcommand{\theequation}{\thesection.\arabic{equation}}

\newcounter{appendice}
\newcommand{\appendice}
{
\setcounter{equation}{0}
\renewcommand{\theequation}{\Alph{appendice}.\arabic{equation}}
\addtocounter{appendice}{1}

{\Large{\bf  Appendix \Alph{appendice}}}
}

\def\thebibliography#1{{\bf REFERENCES\markboth
 {REFERENCES}{REFERENCES}}\list
 {[\arabic{enumi}]}{\settowidth\labelwidth{[#1]}\leftmargin\labelwidth
 \advance\leftmargin\labelsep
 \usecounter{enumi}}
 \def\newblock{\hskip .11em plus .33em minus -.07em}
 \sloppy
 \sfcode`\.=1000\relax}

\def\BI{{\rm 1\!l}}

\begin{document}
\centerline{\LARGE Non-commutative $AdS_2/CFT_1$ duality: }
\vskip .25cm
\centerline{\LARGE  the case of massless scalar fields }
\vskip .5cm

\centerline{A. Pinzul${}^1$\footnote{â€¡
apinzul@unb.br} and A. Stern$^{2}$\footnote{astern@ua.edu}   }

\vskip 0.5cm

\begin{center}
  {${}^1$ Universidade  de Bras\'{\i}lia, Instituto de F\'{\i}sica\\
70910-900, Bras\'{\i}lia, DF, Brasil\\
and\\
International Center of Physics\\
C.P. 04667, Bras\'{\i}lia, DF, Brazil
\\}

\end{center}

\begin{center}
  {${}^2$ Department of Physics, University of Alabama,\\ Tuscaloosa,
Alabama 35487, USA\\}

\end{center}
\vskip 1cm

\normalsize
\centerline{\bf ABSTRACT}
We show how to construct correlators for the $CFT_1$ which is dual to  non-commutative $AdS_2$  ($ncAdS_2$).
We do it explicitly for the example of the massless scalar field  on Euclidean $ncAdS_2$.  $ncAdS_2$   is the quantization of  $AdS_2$ that preserves all the isometries.  It is described in terms of the unitary irreducible representations, more specifically  discrete series representations, of $so(2,1)$.    We  write down  symmetric differential representations for the discrete series, and then map them to functions on  the Moyal-Weyl plane.  The Moyal-Weyl plane has a large distance limit which can be identified with the boundary of $ncAdS_2$.   Killing vectors can be constructed on $ncAdS_2$  which   reduce to the $AdS_2$ Killing vectors near the  boundary.  We  therefore conclude that $ncAdS_2$ is asymptotically $AdS_2$, and so the $AdS/CFT$ correspondence should apply.  For the example of the massless scalar field on Euclidean $ncAdS_2$, the on-shell action, and resulting two-point function for the boundary theory, are computed  to leading order in the noncommutativity parameter.    The computation is nontrivial because nonlocal interactions appear in the Moyal-Weyl description.  Nevertheless, the result is remarkably simple and agrees with that of the commutative scalar field theory, up to a  re-scaling.

\bigskip
\bigskip

\newpage

\section{Introduction}

The \textit{AdS/CFT} correspondence \cite{Maldacena:1997re} has been one of the main themes in theoretical physics for the last 20 years (see, e.g. \cite{Beisert:2010jr} for some recent review). This conjectured correspondence is the explicit realisation of the holographic principle \cite{tHooft:1993dmi,Susskind:1994vu}. In the case of the \textit{AdS/CFT} correspondence this principle is realized in the form of the weak/strong duality between the quantum gravity in the bulk of an asymptotically $AdS$ space and a conformal field theory (\textit{CFT}) on the conformal boundary of this space. The original proposal was made for the case of $AdS_5 \times S^5$ geometry,  in addition to a variety of asymptotically $AdS$ spaces of different dimensions.  A well studied case is that of $AdS_3 / CFT_2$ correspondence. This is due to the fact that the conformal symmetry in two dimensions is infinite dimensional and, as the consequence, the corresponding CFTs are very well studied. It would seem that going one dimension down should simplify things even more. Unfortunately this is not the case.   $AdS_2 / CFT_1$ correspondence\cite{Strominger:1998yg} appears far from being settled. There are several reasons why  this seemingly simple case is more complicated  on both sides of the duality. For example, the geometry of $AdS_2$ is  distinct  from $AdS_n,\,n>2$ because it has two disconnected time-like boundaries.  On the $CFT $ side, there is a realization of $CFT_1$ (which  is actually conformal quantum mechanics rather than field theory),  the de Alfaro-Fubini-Furlan (dAFF) model, which has been known for some time.\cite{deAlfaro:1976vlx}  Although it lacks an $SO(2,1)$-invariant ground state,  it was argued in \cite{Chamon:2011xk} that despite this fact one still can have  correlators consistent with the correspondence. 
Another realization of $CFT_1$ is   matrix quantum mechanics, which is obtained from the dimensional reduction of ten-dimensional super-Yang-Mills theory.\cite{Itzhaki:1998dd}-\cite{Craps:2016cgo} 
Recently a completely different realization of $AdS_2 / CFT_1$ was suggested in \cite{Maldacena:2016hyu,Jensen:2016pah}. There it was conjectured that gravity on (nearly) $AdS_2$ is dual to the so-called Sachdev-Ye-Kitaev models (see references in \cite{Maldacena:2016hyu,Jensen:2016pah}). Though this proposal has attracted  considerable attention, in general, the case of $AdS_2 / CFT_1$ correspondence is still begging for better understanding. In this situation any effort in this direction should be welcome.

In this paper we want to study  aspects of the $AdS_2 / CFT_1$ correspondence in a non-commutative setting, namely when the geometry on the gravity side of the correspondence is replaced by the non-commutative version of  (Euclidean) $AdS_2$. In this regard, two questions naturally  arise:

\noindent 1) Why would one like to make the geometry non-commutative?

\noindent  2) How can we study the non-commutative generalization of the correspondence when, as we mentioned above, even the commutative case is not yet settled?

Concerning the first question, we can argue as follows. There is a general belief (supported by multiple arguments) \cite{Doplicher:1994tu} that the quasiclassical regime of quantum gravity should appear as a quantum field theory on some non-commutative background. In this regard, making the $AdS_2$ space  non-commutative should correspond to the inclusion of some quantum gravitational corrections. Since it is conjectured that the $AdS/CFT$ correspondence is exact even at the quantum level,  it is worthwhile to take non-commutative effects into account  in order  to see explicitly  how (or if) the correspondence applies.  $AdS_2$ is an ideal candidate for examining  non-commutative effects.  This is because it is possible to construct a non-commutative version, $ncAdS_2$, of  $AdS_2$ in such a manner that preserves the isometry group $SO(2,1)$.\cite{Ho:2000fy}-\cite{Chaney}  (This does not mean that the Killing vectors retain their commutative form under the deformation.)  A similar  example of a non-commutative space is the fuzzy sphere $S^2_F$.\cite{Madore:1991bw}-\cite{Ydri:2016dmy} (See  \cite{Alexanian:2000uz} for an explicit  efficient construction for recovering the commutative limit.) Like  $ncAdS_2$, it has the feature of an  undeformed isometry, which proved to be both physically and mathematically useful.  Unlike $S^2_F$, the notion of a boundary can be defined for  $ncAdS_2$, and this is done  purely in terms of
 states of the unitary irreducible representations (UIRR's) of  $SO(2,1)$,  or more generally its universal cover. The non-commutative version of Killing vectors for $AdS_2$ reduce to the commutative form at the boundary.  In this sense,   $ncAdS_2$ can be said to be asymptotically $AdS_2$, and the $AdS/CFT$ correspondence principal should then be applicable.

We have only a  partial answer to the second question. Of course, we will not be able to construct the full correspondence. Instead our goal is more modest: We want to study the perturbative corrections to the correlator functions of operators on the boundary  induced by the bulk-to-boundary and  bulk-to-bulk propagators, and see if they preserve the form  which is compatible with  conformal symmetry. It is possible that the conformal symmetry gets deformed, and this was recently shown in \cite{Gupta:2017xex} where a model of conformal quantum mechanics in $\kappa$-spacetime was considered. This led to  non-commutative corrections to the scaling dimensions. We will see, on the other hand, that such a result does not follow from our construction of  $ncAdS_2$, which is essentially unique when one insists on preserving the isometry group when passing to the noncommutative theory.

We shall assume the usual prescription for the $AdS/CFT$ correspondence, namely, that  the connected correlation functions for  operators ${\cal O}$ spanning the $CFT$ are generated by the on-shell field theory action on the corresponding asymptotically $AdS$ space, and that the boundary values $\phi_0$ of the fields  are sources associated with ${\cal O}$.  In this article we specialize to the case of a single massless scalar field.  This provides a particularly  simple example, in part because of the fact that  solutions to the field equation on $AdS_2$ are regular at the boundary, i.e.,  $|\phi_0| <\infty$.  Moreover, we find that this property is preserved when passing to the noncommutative theory.

Before going to the noncommutative theory, we first briefly recall how the correspondence works for a massless scalar field  $\Phi^{(0)}$ on Euclidean $AdS_2$.
One starts with the action
\beqa  S[\Phi^{(0)}]
&= &\frac 12\int_{ {\mathbb{R}}\times  {\mathbb{R}}_+} dt dz\,\,\Bigl\{( \partial_z\Phi^{(0)} )^2 \,+\,(\partial_t\Phi^{(0)})^2\Bigr\}\;,\label{clmsfa}
\eeqa
where it is convenient to use Fefferman-Graham coordinates, $(z,t)$, $z\ge 0,\,-\infty<t<\infty$,    which we review in section two.  The $AdS_2$ boundary occurs at $z=0$.  Variations $\delta\Phi^{(0)}$ of $\Phi^{(0)}$ in  (\ref{clmsfa}) give
\beqa  \delta S[\Phi^{(0)}]
&= &-\int_{ {\mathbb{R}}\times  {\mathbb{R}}_+} dt dz\,\delta\Phi^{(0)}\,( \partial_z^2+\partial_t^2)\Phi^{(0)} - \int_{ {\mathbb{R}}} dt \,(\partial_z\Phi^{(0)}\,\delta\Phi^{(0)})\Big|_{z=0} \ .
\eeqa
Extremizing the action with Dirichlet boundary conditions yields the field equation
\be \Box\,\Phi^{(0)}= (\partial^2_z + \partial_t^2 )\,\Phi^{(0)}=0 \label{cmtvsffe} \ . \ee
Since the equation is second order we should impose two boundary conditions to obtain a unique solution. Solutions which are everywhere (and in particular at $z\rightarrow\infty$) regular can be expressed in terms of   the boundary value of the field,
 $ \phi_0(t)=\Phi^{(0)}(0,t)$, using the boundary-to-bulk propagator\cite{Witten:1998qj}\footnote{This result is  simple to show in two dimensions: (\ref{frstrdrntsln}) is a solution to the field equation (\ref{cmtvsffe}) since it can be written as $\Phi^{(0)}(z,t)=f(t+iz) + g(t-iz)$, where  $f(t+iz)=\frac i{2\pi}\int_{ {\mathbb{R}}} dt'\,\frac{\phi_0 (t')}{t+iz- t'}$ and $g(t-iz)=f(t+iz)^*$. In the limit $z\rightarrow 0$, the Sokhotski formula gives $\;\frac{1}{t+iz- t'}\rightarrow -i\pi \delta(t-t')+{\cal P}(\frac 1{t-t'})$, where ${\cal P}$ denotes the principal value, and so  $\;\lim_{z\rightarrow 0}\Phi^{(0)}(z,t)=\phi_0(t)$. 
  To see that (\ref{frstrdrntsln}) is regular at $z\rightarrow\infty$ we can write it  as $$\Phi^{(0)}(z,t)= \frac\pi  z\int_{ {\mathbb{R}}} dt'\,\frac{\phi_0 (t')}{1+
 \frac{(t- t')^2}{z^2 }} \;,$$ which for suitable $\phi_0$ tends to zero as $z\rightarrow\infty$.}
 \be\Phi^{(0)}(z,t)= \int_{ {\mathbb{R}}} dt'\, K(z,t;t')\,\phi_0 (t')\;\;,\qquad K(z,t;t')=\frac{z/\pi}{z^2+(t- t')^2}\;\;.\label{frstrdrntsln}  \ee  Denote such solutions by $\Phi_{sol}[\phi_0]$.  They are then substituted back into the action (\ref{clmsfa}), which can also be written as
 \beqa  S[\Phi^{(0)}]
&= &-\frac 12\int_{ {\mathbb{R}}\times  {\mathbb{R}}_+} dt dz\,\Phi^{(0)}\Box\Phi^{(0)}\;-\frac 12\int_{ {\mathbb{R}}} dt \,(\Phi^{(0)}\partial_z\Phi^{(0)})\Big|_{z=0}\;,\label{clmsfasi2}
\eeqa
to obtain the on-shell action.  This leaves  only the boundary term
\beqa  S[\Phi_{sol}[\phi_0]]&=&-\frac 12\int_{ {\mathbb{R}}} dt \;\Phi_{sol}[\phi_0] \; \partial_z\Phi_{sol}[\phi_0] \Big|_{z=0}\label{cmtvBndTurm}\\ &&\cr&=&-\frac{1}{2\pi}\int_{ {\mathbb{R}}} dt \int_{ {\mathbb{R}}} dt'\,\frac{\phi_0(t)\phi_0(t') }{(t-t')^2}\;\;.
\eeqa
In the  $AdS/CFT$ correspondence one identifies $ S[\Phi_{sol}[\phi_0]]$ with the generating functional of the
$n-$point connected correlation functions for the  operators ${\cal O}$ associated with $\phi_0$.  Here, both ${\cal O}$ and  $\phi_0$ are  functions of  only $t$,
 \be <{\cal O}(t_1)\cdots{\cal O}(t_n)>=\frac{\delta^n S[\Phi_{sol}[\phi_0]]}{\delta \phi_0(t_1)\cdots\delta \phi_0(t_n)}\bigg|_{\phi_0=0}\label{adscft} \ .\ee
 So the two-point function in this example is
\be <{\cal O}(t){\cal O}(t ')>\;=\;-\frac{1}{2\pi}\,\frac{1 }{(t-t')^2}\label{2ptfncmtv} \ .\ee

The goal of this article is to repeat the above procedure for scalar fields on Euclidean $ncAdS_2$.
 The  action (\ref{clmsfa}) is replaced by an operator trace.  No additional  terms analogous to the
Gibbons-Hawking-York boundary term\cite{York:1972sj} need to be added to the action for the variational problem to be well defined.  The field equation (\ref{cmtvsffe}) gets replaced by an equation involving infinitely many derivatives in $t$, but still only two derivatives in $z$. Then again only two boundary conditions on a $t-$slice are required to obtain  unique solutions. Regular solutions can be found order by order in the noncommutativity parameter, which can once again be expressed in terms of its boundary values $\phi_0$.  Following an analogous procedure to the above, we obtain the leading order correction to the two-point function (\ref{2ptfncmtv}).

The outline of the article is as follows:  In section two we review Euclidean $AdS_2$ for which we consider two different parametrizations; one are what we call canonical coordinates and the other are  Fefferman-Graham coordinates.  A Poisson bracket is attached to  $AdS_2$ in a manner  consistent with the isometries. The Poisson brackets imply that the time is canonically conjugate to the radial coordinate, which is conventionally interpreted as the energy scale for the boundary $CFT$. In section three we `quantize' the Poisson manifold, and as we indicated previously, we do it in a manner that preserves the  $AdS_2$ isometries.  The result is  $ncAdS_2$, which is described by the UIRR's of  the universal cover of $SU(1,1)$. Of the different nontrivial UIRR's, i.e., the  principal,  supplemental and discrete series, only the discrete series   has a limit back to {\it Euclidean} $AdS_2$, and it is the subject of section four.\footnote{The principal series has a limit to Lorentzian  $AdS_2$,\cite{Jurman:2013ota} while the supplemental series has no continuum limit.} Following  \cite{Groenevelt}, we utilize properties of the generalized Laguerre polynomials to write down a symmetric differential representations for the discrete series.   The differential operators can then be mapped to functions on  the Moyal-Weyl plane, and so one arrives at a convenient Moyal-Weyl description of  $ncAdS_2$.  Furthermore, a boundary can be defined on  the Moyal-Weyl plane  which coincides with the  boundary of $AdS_2$  in the commutative limit.  The Killing vectors for $AdS_2$ have a straightforward analogue in the non-commutative theory and are constructed in section five.   They are realized by infinite order derivative operators on the Moyal-Weyl plane, and as stated above, they  preserve the isometry algebra and reduce to the commutative form near the boundary.  We explore  massless scalar field theory on $ncAdS_2$ in section six.  An explicit expression for the  dynamics of the massless scalar field on $ncAdS_2$ is given.   Although it describes a free scalar field on   $ncAdS_2$, after being mapped to the Moyal-Weyl plane the scalar field picks up nontrivial nonlocal interactions.  Just as with the case of the Killing vectors, the field equation essentially reduces to the commutative equation near the boundary. The field equation can be consistently obtained from the action principle upon imposing Dirichlet boundary conditions, and this is because we find no non-commutative  corrections to the boundary term from variations of the action. The on-shell action, and resulting two-point function for the boundary theory, are computed in section seven to leading order in the noncommutativity parameter.  We find that the results agree with those of the commutative scalar field theory, up to a  rescaling. In appendix we collect some useful results about the Moyal-Weyl star product used in the calculations presented in the sections 5, 6 and 7.

\section{ Euclidean $AdS_2$;  Canonical coordinates versus Fefferman-Graham coordinates}
\setcounter{equation}{0}

$AdS_2$  can be defined in terms of  embedding coordinates $X^\mu$, $\mu=0,1,2$, along with a scale parameter $\ell_0$. In the case of the Euclidean version of $AdS_2$, $X^\mu$ span three-dimensional Minkowski space with invariant interval  $ds^2=dX^\mu dX_\mu$, where indices raised and lowered using the ambient metric tensor  $\eta={\rm diag}(1,1,-1)$.  $AdS_2$ is defined by the constraint
\be X^\mu X_\mu
=-\ell_0^2\;,\label{Ucldads}\ee
and $\ell_0^2>0$.
 The constraint describes a double-sheeted hyperboloid embedded in $3$d Minkowski space-time. $AdS_2$ has three Killing vectors $K^\mu$ which generate the $SO(2,1)$ isometry group and get identified with the generators of the global conformal symmetry on the boundary.  Thus
\be [K^\mu, K^\nu]=\epsilon^{\mu\nu\rho}K_\rho\;\;.\label{klngvctr}\ee
 Our convention for the Levi-Civita symbol is $\epsilon^{012}=1$. The action of the Killing vectors on the embedding coordinates is
\be (K^\mu X^\nu)=\epsilon^{\mu\nu\rho}X_\rho\;\;.\label{klngvctronxmu}\ee

For the purpose of quantization we  attach a Poisson bracket  to the $AdS$ manifold.  In two dimensions one can introduce a Poisson bracket  which respects the  isometry group, and therefore, the global conformal symmetry at the boundary.  Expressed in terms of the embedding coordinates it is
 \be \{ X^\mu, X^\nu\}=\epsilon^{\mu\nu\rho}X_\rho\;\;.\label{su11pbs}\ee
In comparing with  (\ref{klngvctronxmu}), the action of Killing vectors on  functions on  $AdS_2$ can the be written as $K^\mu =\{X^\mu,\cdot\}$.

Two choices of coordinates on the surface are useful for us.
One choice  is $\{(x,y),\, -\infty<x,y<\infty\}$, defined by
\be  X^0=-y\quad ,\quad  X^1=- \frac {1}{2\ell_0} \,e^{-x}y^2+{\ell_0}\,\sinh{x}\quad ,\quad
 X^2=- \frac {1}{2\ell_0} \,e^{-x}y^2 -{\ell_0}\,\cosh{x}\;\ .\label{smtrcdfrsplonpssp}\ee
 It covers a single hyperboloid ($X^2<0$).
 In terms of these coordinates, the  metric tensor induced on the surface from the  $3$d Minkowski  metric is given by
  \be   ds^2=\ell_0^2\, dx^2 + (dy - y dx)^2 \;,\label{mtrcinxy}\ee
while the three Killing vectors are
\be K^0=\partial_x\quad ,\quad K^1=  \frac 1{\ell_0} e^{-x}y\,\partial_x\,-\, X^2\,\partial_y\quad ,\quad K^2=  \frac 1{\ell_0} e^{-x}y\,\partial_x\,-\, X^1\,\partial_y\;\;.\label{Kmuincancds}\ee
The coordinates $(x,y)$ have the feature that they are canonically conjugate.  That is, upon assuming that
\be \{x,y\}=1\label{cnclpbsxy}\;,\ee
and  using (\ref{smtrcdfrsplonpssp}),  we recover the invariant Poisson brackets (\ref{su11pbs}). For this reason refer to $(x,y)$ as canonical coordinates.

A more familiar parametrization of the hyperboloid is given by the Fefferman-Graham coordinates $(z,t)$
\be z=e^{-x}\quad ,\qquad t=\frac 1{\ell_0}e^{-x} y \ .\label{mp2pncra}\ee
Whereas the canonical coordinates span $ {\mathbb{R}}^2$, $(z,t)$ cover the half-plane, $z\ge 0,\,-\infty<t<\infty$.  $r=z^{-1}$ can be regarded as a radial variable.  It can be expressed linearly in terms of the embedding coordinates,
\be r=z^{-1}=\frac 1 {\ell_0}(X^1-X^2)\ .  \label{rdlvrble} \ee  The  $AdS_2$ boundary is   the open curve at $z=0$ or $r\rightarrow\infty$.  In terms of the canonical coordinates, the boundary corresponds to both $x$ and $y$ going to infinity, with $e^{-x} y$ finite.   The metric tensor when expressed in Fefferman-Graham coordinates is given by
\be   ds^2=\frac{\ell_0^2}{z^2}\;\Bigl( dz^2+ dt^2\Bigr) \,\;,\ee
 and the Killing vectors take the form
\be K^-=-\partial_t\quad ,\quad K^0=-t\partial_t-z\partial_z\quad ,\quad K^+=(z^2- t ^2)
\,\partial_t -2 z t \,\partial_z \;,\label{KlngfG}
\ee  where  $K^\pm=K^2\pm K^1$.
We see that in the limit $z\rightarrow 0$, one recovers  the standard form for the global conformal symmetry generators on the boundary
\be K^-\rightarrow -\partial_t \quad ,\quad K^0 \rightarrow -t\partial_t\quad ,\quad K^+\rightarrow - t^2\partial_t\ . \label{kmuFGcrds}
\;\ee
They generate, respectively,  translations, dilatations and special conformal transformations on the boundary.
In terms of  the   Fefferman-Graham coordinates the Poisson bracket  which yields the $so(2,1)$ Lie algebra (\ref{su11pbs})  is
\be\{t,z\}=\frac 1{\ell_0}z^2 \ .\label{poissontau}\ee
From (\ref{rdlvrble}) it also follows that
\be \{r,t\} =\frac 1{\ell_0}\;\;.  \label{rtcnclycngt}\ee
In the $AdS/CFT$ correspondence the radial variable is often regarded as the energy scale for the boundary
$CFT$, and so it is reasonable to find that it is canonically conjugate to the time $t$. Note that in passing to the quantum theory we cannot simply replace the variables $r$ and $t$ with self-adjoint operators since $r$ is only defined on the half-line.  An alternative way to proceed to the quantum theory will be given in the following section.

We note that if do yet another change of coordinates from $(z,t)$ to complex coordinates  $\zeta=t+iz$ and  $\bar\zeta=t-iz$,  the Killing vectors become
$   K^{\mu} = L^\mu + \bar L^\mu \;,$ where $ L^\mu $, along with their complex conjugates $\bar L^\mu $, are the standard global conformal symmetry generators on the complex plane
\be L^-= -\partial_\zeta \quad ,\quad L^{0}= -\zeta\partial_\zeta \quad ,\quad L^{+}= -\zeta^2\partial_\zeta  \ . \ee The Poisson bracket (\ref{poissontau}) written in terms of $\zeta$ and $\bar{\zeta}$ will become
\be \{ \zeta , \bar{\zeta}\} = \frac{i}{2\ell_0}(\zeta - \bar{\zeta})^2 \ .\ee This bracket can be quantized using the methods of \cite{Alexanian:2000uz} to produce a star-product written  directly in terms of the Fefferman-Graham coordinates (it is expected to be highly non-trivial). In this paper we will not follow this line.

\section{ $ncAdS_2$}
\setcounter{equation}{0}

There is a straightforward quantization of the Poisson manifold defined in the previous section, and the result is  $ncAdS_2$.\cite{Ho:2000fy}-\cite{Chaney}
The first step is to replace the three embedding coordinates $ X^\mu$ by hermitean operators  $ \hat X^\mu$.  The analogue of the constraint (\ref{Ucldads}) in this setting is
\be  \hat X^\mu \hat X_\mu=-\ell^2\BI \label{Ucldadscasimir}\;, \ee where $\BI$ is the identity
and $\ell^2>0$ in the  Euclidean version of  $ncAdS_2$. Furthermore, following the usual quantization procedure, the Poisson brackets (\ref{su11pbs}) are promoted  to
 commutation relations
\be [\hat X^\mu,\hat X^\nu]=i\alpha\,\epsilon^{\mu\nu\rho}\hat X_\rho\;\;.\label{adstoocrs}\ee  $\alpha $ and $\ell$ are two real parameters with units of length. (\ref{Ucldadscasimir}) and (\ref{adstoocrs}) define $ncAdS_2$, which is a solution to certain matrix models, which we describe below.\cite{Jurman:2013ota}-\cite{Chaney}
 The commutation relations (\ref{adstoocrs}) define the $ so(2,1)$ algebra, while (\ref{Ucldadscasimir}) fixes a value of the $ so(2,1)$ Casimir operator.    Analogous to (\ref{rdlvrble}), one can  construct an operator analogue of the radial coordinate from the hermitian operators $\hat X^\mu$
 \be \hat r=\frac 1 {\ell}(\hat X^1-\hat X^2)\; \label{rdlpsop} \;\;.\ee
 We obtain the spectrum and  eigenfunctions of this operator in section four.

   Both (\ref{Ucldadscasimir}) and (\ref{adstoocrs}) are preserved under the  action of $SO(2,1)$, $\hat X^\mu\rightarrow R^\mu_{\;\;\nu} \hat X^\nu$, where $R$ is a $SO(2,1)$ matrix.  This is the analogue of  isometry transformations on $AdS_2$.  We shall construct the non-commutative analogues of the Killing vectors
(\ref{Kmuincancds}) and (\ref{KlngfG}) which generate such transformations in section five. In addition to the $SO(2,1)$ symmetry, the equations (\ref{Ucldadscasimir}) and (\ref{adstoocrs}) are invariant under unitary transformations   $\hat X^\mu\rightarrow \hat U\hat X^\mu \hat U^\dagger$, where $\hat U$ is a unitary operator.

To show that (\ref{Ucldadscasimir}) and (\ref{adstoocrs}) can be obtained from matrix models\cite{Jurman:2013ota}-\cite{Chaney}   one can introduce
  three infinite-dimensional hermitean matrices $Y^\mu$, $\mu=0,1,2$, with an action $S_M(Y)$ consisting of two terms
\be S_M(Y)={\rm Tr}\Bigl(-\frac 14 [Y_\mu, Y_\nu] [Y^\mu,Y^\nu]  {-}\frac 23 i \alpha \epsilon_{\mu\nu\lambda}Y^\mu Y^\nu Y^\lambda\Bigr)\;,\label{mmactn}\ee  where Tr denotes a matrix trace and we again assume  the ambient metric $\eta_{\mu\nu}=$diag$(1,1,-1)$. Dynamics can be defined by adapting a variational principle to this system.  Extremizing $S_M$ with respect to variations in $Y_\mu$ leads to
\be [ [Y^\mu,Y^\nu],Y_\nu] {-}i\alpha \epsilon^{\mu\nu\lambda}[Y_\nu,Y_\lambda] =0 \ .\label{mmeqofmot} \ee
They are clearly solved by setting $Y^\mu=\hat X^\mu$. Like $ncAdS_2$, the matrix equations (\ref{mmeqofmot}) possess $SO(2,1)$ invariance, as well as invariance under unitrary transformations (where $\hat U$ now denotes an infinite dimensional unitary matrix).  The matrix equations have an additional translational symmetry, $Y^\mu\rightarrow Y^\mu+v^\mu\BI$, where $\BI$ is the unit matrix and $v^\mu$ are real, which is broken by the $ncAdS_2$ solution.  Other matrix models have $ncAdS_2$ solutions.  For example, one can add a mass term,
Tr$\,Y^\mu Y_\mu$, to (\ref{mmactn}), and consequently a linear term to the equations of motion (\ref{mmeqofmot}),  as was done in \cite{Chaney}.  This term explicitly breaks the translation symmetry.

To recover $AdS_2$ from $ncAdS_2$, we  need to define the commutative limit.  It is $(\alpha,\ell)\rightarrow (0,\ell_0)$.  In that limit, (\ref{Ucldadscasimir}), (\ref{adstoocrs}) and (\ref{rdlpsop}) go to  (\ref{Ucldads}), (\ref{su11pbs}) and (\ref{rdlvrble}), respectively. Here $\alpha$ plays the role of $\hbar$.  It will also be necessary to define the notion of a boundary limit in the non-commutative theory.  A natural choice for this  is that the limit of the expectation value of $\hat r$ becomes large.
This limit can be made more precise upon specifying the Hilbert space of the system, which we do below.

The states of  $ncAdS_2 $ belong to
 unitary irreducible  representations of $SU(1,1)$ [More precisely, it is the universal cover of the groups because we shall only be concerned with representations of the algebra (\ref{adstoocrs})], which are the principal, supplemental and discrete series representations.  They  are in general labeled by two parameters, which we denote by $\epsilon_0$ and $k$. One can take a basis  in a given representation to be eigenvectors   $\{|\epsilon_0,k,m>$, $m=$integer$\}$ of $\hat X^2$.  The integer $m$ is raised and lowered by $\hat X_+=\hat X^1+ i \hat X^0 $ and $\hat X_-=\hat X^1- i\hat  X^0 $, respectively.  Thus
\beqa
\hat X_+|\epsilon_0,k,m>&=&-\alpha\, c_m  \,|\epsilon_0,k,m+1> \ ,\label{oneone1}
\\ &&\cr
\hat X_-|\epsilon_0,k,m>&=&-\alpha\,  c_{m-1}\,|\epsilon_0,k,m-1>\;,\label{rsnglwrng}\\ &&\cr
\hat X^2|\epsilon_0,k,m>&=&-\alpha\, (\epsilon_0 +m)\,|\epsilon_0,k,m>\label{x1ignvlu}\;,\label{csubm}\eeqa
where   the coefficient $c_m $ is \be c_m = \sqrt{(k+\epsilon_0+m+1)(\epsilon_0-k+m)}\ee which  ensures that the basis vectors are orthonormal $<\epsilon_0,k,m|\epsilon_0,k,m'>=\delta_{m,m'}$.
For any irreducible representation the Casimir operator is fixed by
\beqa
 \hat X^\mu \hat X_\mu|\epsilon_0,k,m>&=&-\alpha^2\,k(k+1)\,|\epsilon_0,k,m>\;\;,\label{csmrkp1}\eeqa
Upon  comparing with (\ref{Ucldadscasimir}) we then get, \be \Bigl(k+\frac 12\Bigr)^2=\frac 14+\frac{\ell^2}{\alpha^2}\ .\label{klalfa}\ee

 We note that the right hand side of (\ref{klalfa}) diverges in the commutative limit.  Therefore the commutative limit corresponds to the limit of representations with $k\rightarrow \pm\infty$.
From (\ref{rdlpsop}), the expectation value of the radial position vector $\hat r$  for any eigenvector  $|\epsilon_0,k,m>$ is $\frac\alpha {\ell} (\epsilon_0+m)$.  Since the expectation value grows with $m$ one can associate the boundary of  $ncAdS_2$ with $m\rightarrow \infty$.

As is well known there are three different types of nontrivial unitary irreducible representations of  $SU(1,1)$: the principal,  supplemental and discrete series representations. These series are distinguished by their allowed values for $k$. The principal series representation has   $k=-\frac 12- i \rho$, where $\rho$ is real. This means that Casimir in (\ref{csmrkp1}) is positive and $\ell$ in (\ref{Ucldadscasimir}) is imaginary. This corresponds to the Lorentzian version of  $ncAdS_2$ which we are not considering here.  Moreover, the limit $\rho\rightarrow\infty$, $\alpha\rightarrow 0$ yields  Lorentzian $AdS_2$, which was pointed out in \cite{Jurman:2013ota}.  As our interest is  in recovering Euclidean $AdS_2$,  we do not examine the principal series.  The  supplemental series
has $k$ real, but restricted to $-\frac 12 <k<0$.  The  Casimir in (\ref{csmrkp1}) is again positive and $\ell$ is imaginary. But since we cannot take the limit $k\rightarrow\infty$ in this case, the  supplemental series has no commutative limit. We can say that this case describes purely quantum Lorentzian $ncAdS_2$.\footnote{Also, from (\ref{klalfa}) it is clear that $|\ell|< \frac{\alpha}{2}$. So this means that this space is an extremely quantum object without any commutative limit.} For these reasons we shall also not consider  the  supplemental series.  We  note that
$m$ ranges over all positive and negative integers for the principal and supplemental series.  This means that the expectation values of $\hat r$ are not restricted to being positive.  Moreover,  $m\rightarrow\infty $ and $m\rightarrow-\infty $ are permissible limits of the states, which can be associated with two boundaries for the non-commutative version of Lorentzian $AdS_2$.

In the case of two discrete series representations,  $D^\pm(k)$, $k$ can be an arbitrary negative number.\footnote{If one were  to specialize to UIRR's  of  $SU(1,1)$, rather than its universal covering group,  then one can show that $k$ is restricted to the negative half-integer numbers.\cite{Bargmann:1948ck,Barut:1965,Balachandran:1969nc,Balnotes} But since here we are  only concerned with representations of the algebra (\ref{adstoocrs}), this restriction  is not necessary.}
Therefore the Casimir in (\ref{csmrkp1}) is negative (and hence $\ell$ is real) for  $k<-1$, and so these representations describe Euclidean   $ncAdS_2$.\footnote{The non-negative Casimir for $k\in [-1,0)$ describes an extremely quantum space, so it does not make much sense to say that for these values of $k$ we have a Lorenzian $ncAdS_2$. In any case, we are interested in the quasiclassical regime, i.e. when $k$ is large.}  Moreover, the limit that $k$ goes to either $+\infty$ or $-\infty$ exists, so the discrete series has a limit to Euclidean  $AdS_2$. $m$ takes on only positive integers (including zero)  for the discrete series representation $D^+(k)$, and negative integers for $D^-(k)$, defining two distinct noncommutative analogues  of the hyperboloids of Euclidean $AdS_2$.

\section{Discrete series representations}
\setcounter{equation}{0}
Here following  \cite{Groenevelt}, we utilize properties of the generalized Laguerre polynomials to write down a symmetric differential representations of $ \hat X^\mu$ for the discrete series representations  $D^+(k)$ and  $D^-(k)$.
We do this by obtaining eigenstates of the radial coordinate operator $\hat r$ in (\ref{rdlpsop}).

We begin with  $D^+(k)$.  Here $\epsilon_0=-k$ is a positive number.  These representations have a lowest state $|-k,k,0>$, which from (\ref{rsnglwrng}) is annihilated by $\hat X_-$.  For brevity we denote this state by $|k,0>$, and all other states  in the $\hat X^2$ eigenbasis by  $|k,m>$, $m\;=$ positive integer.  Next denote the eigenvector of  the radial position operator (\ref{rdlpsop})  by $\widetilde{|r,k>_+}\in  D^+(k)$, and with some abuse of notation, we   call $r$ the eigenvalue,
\be \hat r \widetilde{|r,k>_+}=  r \widetilde{|r,k>_+}\ ,\ee
$ \widetilde{|r,k>_+}$ can be expanded in the $\hat X^2$ eigenbasis,
\be\widetilde{|r,k>_+}\,=\,\sum_{m=0}^\infty\,\psi_{k,m}^+(r)\:|k,m>\;\; .\ee
Recursion relations for the coefficients  $\psi_{k,m}^+(r)$ follow from the definition of $\hat r$, (\ref{rdlpsop}), along with  (\ref{oneone1}) and (\ref{rsnglwrng}),
\be
 \sqrt{(m+1)(m-2k)}\, \psi_{k,m+1}^+(r)+ \sqrt{m(m-1-2k)}\, \psi_{k,m-1}^+(r)+2\Bigl (k-m +\frac {\ell}\alpha r\Bigr)\psi_{k,m}^+(r)=0\;,
\label{dblinftJcbee}\ee which is also valid for $m=0$ since then the second term vanishes, and so all coefficients are determined from $\psi_{k,0}^+(r)$.  The recursion relations  (\ref{dblinftJcbee}) agree with those of the  generalized Laguerre polynomials  $L^{(\gamma)}_m$, $m$ being a non-negative integer, upon setting
\be\psi_{k,m}^+(r)= \sqrt{\frac{m!}{(m-2k-1)!}}\,\;L^{(-2k-1)}_m\Bigl(\frac{2\ell r}\alpha\Bigr) \;\; .\label{psiinglps}\ee
The domain for  $L^{(\gamma)}_m$ is the half-line, and so just as in the commutative theory,  $r\ge 0$.
A single  boundary occurs in this case, corresponding to $r\rightarrow\infty$.
The dominant  polynomials near the boundary have large  $m$, which is consistent with the previous result that   the expectation value of $\hat r$ grows with $m$.

The  generalized Laguerre polynomials obey the differential equation
\be \zeta \frac{d^2}{d\zeta^2} L^{(\gamma)}_m(\zeta)+(\gamma+1-\zeta) \frac{d}{d\zeta}L^{(\gamma)}_m(\zeta)+mL^{(\gamma)}_m(\zeta)=0\;,\ee
and the orthogonality conditions\footnote{This is defined only for $\gamma > -1$ (to avoid the logarithmic divergence at $\zeta =0$), which is satisfied in our case, $k<-1$.}
\be \int_0^\infty d\zeta \,\zeta^\gamma e^{-\zeta}\,L^{(\gamma)}_m(\zeta)\,L^{(\gamma)}_n(\zeta)=\frac 1{n!}\Gamma(n+\gamma +1)\,\delta_{n,m}\ .\ee
Upon writing $\psi_{k,m}^+(r)=\Bigl(\frac{2\ell}\alpha\Bigr)^{k}e^{\frac{\ell}\alpha r}\, r^{k+\frac 12}\,u_{k,m}^+(r)$ and using (\ref{psiinglps}), these two relations can be expressed as
\beqa -\frac \alpha{2\ell} \biggr(\;  \frac d{dr} r\frac d{dr}- \frac{(k+\frac 12)^2}{ r}   -\frac{\ell^2}{\alpha^2}\, r\biggl)u_{k,m}^+(r)& =&(m-k)\; u_{k,m}^+(r)\label{4twofive} \ , \\&&\cr \int_0^\infty dr\,u_{k,m}^+(r)\,u_{k,n}^+(r) &=&\delta_{n,m}\;,\eeqa respectively.
In comparing (\ref{4twofive}) with the eigenvalue equation   (\ref{x1ignvlu}), we get  a symmetric differential representation $\pi^{k}$ of $\hat X^2$ on $L^2(R_+, dr)$
\be  \pi^{k}(\hat X^2)=  \frac {\alpha^2}{2\ell} \biggr(\;  \frac d{dr} r\frac d{dr}- \frac{(k+\frac 12)^2}{ r}   -\frac{\ell^2}{\alpha^2}\, r\biggl)\label{sdrfrx1}\ee
The corresponding   differential representations for the remaining  $ncAdS_2$  operators $\hat X^0$ and $\hat X^1$ are obtained using $\pi^{k}([\hat r,\hat X^2]) =\frac {i  \alpha}\ell \;\pi^{k}(\hat X^0)$ to get the former  and then $\pi^{k}([\hat X^0,\hat X^2]) =-{i  \alpha}\;\pi^{k}(X^1)$ to get  the latter.  The results are
\beqa  \pi^{k}(\hat X^0)&=&i\alpha \Bigl( r \frac d{dr}+\frac 12\Bigr) \ ,\label{pikxz}
\\ &&\cr   \pi^{k}(\hat X^1)&=&  \frac {\alpha^2}{2\ell} \biggr(\;  \frac d{dr} r\frac d{dr}- \frac{(k+\frac 12)^2}{ r}  +\frac{\ell^2}{\alpha^2}\, r\biggl)\ .\label{smtrcdfrspl}\eeqa
As the consistency check, note that from (\ref{rdlpsop}), (\ref{sdrfrx1}) and (\ref{smtrcdfrspl}) follows that $\hat{r}$ is really diagonal in this representation, $\pi^k (\hat{r}) = r$.

For the discrete series $D^-(k)$, $\epsilon_0=k$ and $m$ are negative integers including zero.  The  radial  eigenvector is
\be\widetilde{|r,k>_-}\,=\,\sum_{m=0}^{-\infty}\,\psi_{k,m}^-(r)\:|k,m>\;,\ee
\be\psi_{k,m}^-(r)=(-1)^m \sqrt{\frac{(-m)!}{(-m-2k-1)!}}\;L^{(-2k-1)}_{-m}\Bigl(-\frac{2\ell r}\alpha\Bigr) \;,\ee
which now is defined only for $r\le 0$.  The boundary now is at $r\rightarrow-\infty$ where the polynomials $L^{(\gamma)}_m(\zeta)$ with large negative $m$ dominate.  The above analysis can be repeated for  $D^-(k)$  to obtain expression for the symmetric differential representation of the $su(1,1)$ basis.  The results are again given by  (\ref{sdrfrx1}) and (\ref{smtrcdfrspl}), now acting on  functions spanned by $\{u_{k,m}^-(r),\; r\le 0\}$, which are defined by $\psi_{k,m}^-(r) =\Bigl(\frac{2\ell}\alpha\Bigr)^{k}e^{-\frac{\ell}\alpha r}(-r) ^{k+\frac 12}\;u_{k,m}^-(r)$.

 The  linear operators in (\ref{sdrfrx1})-(\ref{smtrcdfrspl}) act on  $L_2(R_+,dr)$.  Denote the space of square-integrable space of functions on $R_+$ by $\{\psi(r)\}$.  It is convenient to replace $r$ by $x=\log r$ and replace  (\ref{sdrfrx1})-(\ref{smtrcdfrspl}) by linear operators $\tilde \pi^k(\hat X^\mu)$ that on act $L_2(R,dx)$,  spanned by   $\{f(x)= e^{x/2}\,\psi(e^x)\}$.  The result can be expressed in terms of self-adjoint operators $\hat x$ and $\hat y$ on  $L_2(R,dx)$, where $\hat x$ has a trivial action on functions, $\hat x f(x)=x f(x)$, and $\hat y$ is the self-adjoint differential operator $\hat y=-i\alpha\partial_x$.  Then $\hat x$  and $\hat y$   satisfy the Heisenberg commutation relation
\be [\hat x,\hat y]=i\alpha \BI\label{hsnbrgcr}\;,\ee $\BI$ being the identity.
For $\tilde \pi^k(\hat X^\mu)$ we get
\beqa  \tilde\pi^{k}(\hat X^0)&=&-\hat y \ ,
\cr &&\cr   \tilde \pi^{k}(\hat X^1)&=&- \frac {1}{2\ell} \,\hat y\,e^{-\hat x}\hat y -\frac{\alpha^2}{2\ell}k (k+1)\,e^{-\hat x}  +\frac{\ell}{2}\, e^{\hat x} \ ,
\cr &&\cr   \tilde \pi^{k}(\hat X^2)&=&- \frac {1}{2\ell} \,\hat y\,e^{-\hat x}\hat y -\frac{\alpha^2}{2\ell}k (k+1)\,e^{-\hat x}  -\frac{\ell}{2}\, e^{\hat x}\ .\label{smtrcdfrsplonR}\eeqa

Since $\hat x$ and  $\hat y$ satisfy  (\ref{hsnbrgcr}),  any function $\hat {\cal F}(\hat x,\hat y)$ can be mapped to function $ {\cal F}( x, y)$, called symbol, on the Moyal-Weyl plane, which we take to be spanned by commuting variables $x$ and $y$. Then $x$ and $y$ are the symbols  of $\hat x$ and $\hat y$, respectively. [Here we are identifying coordinates of the Moyal-Weyl plane with the canonical coordinates of section two. This is justified by the fact that they coincide in the commutative limit, as we show below.] The product  $[\hat {\cal F}\hat{\cal G}](\hat x,\hat y)= \hat {\cal F}(\hat x,\hat y) \hat{\cal G}(\hat x,\hat y)$ of any two functions of $\hat x$ and  $\hat y$ is
mapped to the Moyal-Weyl star product
 $ [ {\cal F}\star  {\cal G} ](x,y) $, which is written down explicitly in (\ref{dffstrprd}) in the appendix.
We denote the symbols of $\tilde \pi^k(\hat X^\mu)$ by ${\cal X}^\mu$.  Then from (\ref{smtrcdfrsplonR}),
\beqa   {\cal X}^0 &=&- y \ .
\cr &&\cr  {\cal X}^1 &=&- \frac {1}{2\ell} \, y\star  e^{-x}\star y -\frac{\alpha^2}{2\ell}k (k+1)\,e^{- x}   +\frac{\ell}{2}\, e^{ x}\ .
\cr &&\cr  {\cal X}^2 &=&- \frac {1}{2\ell} \, y\star e^{-x}\star y -\frac{\alpha^2}{2\ell}k (k+1)\,e^{- x}   -\frac{\ell}{2}\, e^{ x}\ .\label{smnlsofXmu}\eeqa
These are the analogues of the embedding coordinates $X^\mu$.  They do not satisfy the $AdS_2$ constraint (\ref{Ucldads}) using the point-wise product.
Rather using the star-product (\ref{dffstrprd}), they realize  the defining relations (\ref{Ucldadscasimir}) and (\ref{adstoocrs}) for  $ncAdS_2$ on the Moyal plane
\beqa  {\cal X}^\mu  \star {\cal X}_{ \mu}&=& - \ell^2 \ ,\\
&&\cr  [ {\cal X} ^\mu, {\cal X} ^\nu]_\star &=&i\alpha\,\epsilon^{\mu\nu\rho} {\cal X}_{\rho}\;,\;\eeqa
where  $ [{\cal F}, {\cal G} ]_\star={\cal F}\star {\cal G} -{\cal G}\star {\cal F}$ is the star commutator of any two functions $ {\cal F}( x, y)$ and ${\cal G}( x, y)$ on the  Moyal-Weyl plane, and we have used (\ref{klalfa}). In the commutative limit $\alpha\rightarrow 0$, the star product reduces to the point-wise product, and the leading term in the star commutator is  $ [{\cal F}, {\cal G} ]_\star\rightarrow i\alpha \{{\cal F}, {\cal G}\}$, where $\{\,,\, \}$ denotes  the Poisson bracket defined using (\ref{cnclpbsxy}).  Thus $x$ and $y$ reduce to the canonical coordinates of section two.  Moreover, using (\ref{klalfa}) one can show that ${\cal X} ^\mu$ reduce to the $AdS_2$ embedding coordinates $X^\mu$, eq. (\ref{smtrcdfrsplonpssp}), in the commutative limit.

\section{Killing vectors on   $ncAdS_2$}
\setcounter{equation}{0}

From section two, isometry transformations  on  $AdS_2$ can be obtained by taking  Poisson brackets with $X^\mu$.  Given a function $\Phi$ on  $AdS_2$ an infinitesimal variation of $\Phi$  induced by the action of the $SO(2,1)$ isometry group is
\be \delta \Phi=\epsilon_\mu\,(K^\mu\Phi) =\epsilon_\mu\,\{X^\mu,\Phi\} \;,\ee
where $K^\mu$ are the Killing vectors on  $AdS_2$ and $\epsilon_\mu$ are infinitesimal parameters.
There is a natural generalization to $SO(2,1)$ isometry transformations on  $ncAdS_2$, and hence to  Killing vectors $\hat K^\mu$ on   $ncAdS_2$. If $\hat\Phi$  is a function on  $ncAdS_2$ its infinitesimal variation  $\delta_{nc}\hat \Phi$  induced by the action of  $SO(2,1)$ is
 \be \delta_{nc} \hat \Phi=\epsilon_\mu\,(\hat K^\mu\hat \Phi) = i \epsilon_\mu\,[\hat X^\mu,\hat \Phi] \ .\ee

Alternatively, it can be mapped to infinitesimal transformations on the  Moyal-Weyl plane.
If we call $\Phi$ the symbol of $\hat \Phi$ and $K^\mu_\star \Phi$ the symbol of $\hat K^\mu\hat \Phi$ then
\be \delta_{nc} \Phi=\epsilon_\mu\,( K^\mu_\star\Phi)=i\epsilon_\mu\,[{\cal X}^\mu ,\Phi]_\star \ .\label{su11trnsfrm}\ee
 Using (\ref{dffstrprd}) and the expressions (\ref{smnlsofXmu}) for ${\cal X}^\mu$ we get
\be \delta_{nc}\Phi = \alpha \epsilon_0 \partial_x\Phi +\frac {i\epsilon_+} {2\ell}\,[ y\star e^{- x}\star y,\Phi]_\star  +\frac {i\epsilon_+\alpha^2} {2\ell}k(k+1)\,[e^{-x},\Phi]_\star+\frac{i\epsilon_- \ell} {2}\,[e^x,\Phi]_\star\;,\label{tre5fiv}\ee
where $\epsilon_\pm=\epsilon_2\pm \epsilon_1$.
The variation can be explicitly computed with the help of the identities (\ref{MWidntes}) in the appendix. One gets
\beqa [e^{\pm x},\Phi]_\star &=&\pm i\alpha\, e^{\pm x} \Delta_y\Phi \ ,\cr &&\cr
[ y\star e^{- x}\star y,\Phi]_\star &=&-i\alpha\, e^{-x}\,\biggl(y^2 \Delta_y+2 y\partial_x S_y+\frac{\alpha^2}4(1-\partial_x^2)\Delta_y\biggr)\,\Phi
\;,\label{strcmtrsFi}\eeqa
where \beqa \Delta_y\Phi(x,y)&=&\frac{\Phi\Bigl(x,y+\frac {i\alpha}2\Bigr)-\Phi\Bigl(x,y-\frac {i\alpha}2\Bigr)}{i\alpha}\;=\; \frac 2\alpha\sin\Big(\frac{\alpha}2\partial_y\Bigr)\Phi(x,y)\ ,\cr &&\cr  S_y\Phi(x,y)&=&\frac{\Phi\Bigl(x,y+\frac {i\alpha}2\Bigr)+\Phi\Bigl(x,y-\frac {i\alpha}2\Bigr)}{2} \;=\; \cos\Big(\frac{\alpha}2\partial_y\Bigr)\Phi(x,y)\ .\label{dfDltaS}\eeqa
The non-commutative variation  can then be written as
$\delta_{nc}\Phi=\frac  \alpha 2 \, \Bigl({\epsilon_-}   K_\star^{-}+2\epsilon_0 K_\star^{0}+{\epsilon_+}  K_\star^{+}\Bigr) \Phi $,
where the non-commutative analogues of the $AdS_2$ Killing vectors are
\beqa K_\star^-&=&-{\ell} \, e^{ x} \Delta_y\qquad\ ,\qquad\quad K_\star^0\;=\;\partial_x \ ,\cr &&\cr K_\star^+&=& \frac {e^{-x}} {\ell}\,\Biggl(2 y\,\partial_x S_y+\biggl(y^2+{\ell^2}+ \frac {\alpha^2}4(1-\partial_x^2)\biggr)\, \Delta_y
\Biggr)\ .\label{Mnxndy}\eeqa
 By construction $K^\mu_\star$  satisfy the $so(2,1)$ Lie algebra commutation relations  $[K_\star^\mu, K_\star^\nu]=\epsilon^{\mu\nu\rho}K_{\star\rho}$, where $K^2_\star =\frac 12 (K^+_\star +K^-_\star)$ and $K^1_\star =\frac 12 (K^+_\star -K^-_\star)$. $K^0_\star$ agrees with its commutative analogue $K^0$, while  $K^1_\star$ and  $K^2_\star$ are deformations of  $K^1$ and  $K^2$, (\ref{Kmuincancds}), containing  infinite  order polynomials in $\partial_y$.
In the commutative limit  $(\alpha,\ell)\rightarrow (0,\ell_0)$,  $\;\Delta_y$ approaches a derivative operator $\partial_y$ and $S_y$ approaches the identity.  It follows that we recover the $AdS_2$ Killing vectors in the commutative limit,  $K^\mu_\star\rightarrow K^\mu$ as $\alpha\rightarrow 0$.

The non-commutative analogues of the Killing vectors
 can be re-expressed in  Fefferman-Graham coordinates (\ref{mp2pncra}) by replacing the action of  $\Delta_y$ and $S_y$ on the fields by
 \beqa \Delta_t\Phi(z,t)&=&\frac{\Phi\Bigl(z,t+\frac {i\alpha z}{2\ell}\Bigr)-\Phi\Bigl(z,t-\frac {i\alpha z}{2\ell}\Bigr)}{i\alpha}\;=\;\frac 2\alpha\sin\Bigl( \frac{\alpha z}{2\ell}\partial_t\Bigr)\,\Phi(z,t) \ ,\cr &&\cr  S_t\Phi(z,t)&=&\frac{\Phi\Bigl(z,t+\frac {i\alpha z}{2\ell}\Bigr)+\Phi\Bigl(z,t-\frac {i\alpha z}{2\ell}\Bigr)}{2}\;=\;\cos\Bigl( \frac{\alpha z}{2\ell}\partial_t\Bigr)\,\Phi(z,t)\eeqa
respectively. Then
\beqa  K_\star^-&=&-\frac{\ell}z \, \Delta_t\qquad\ ,\qquad\quad K_\star^0\;=\;-t\partial_t-z\partial_z\ ,\cr &&\cr  K_\star^+&=&- 2t\,(t\partial_t +z\partial_z) S_t+ \frac {\ell} {z}\,\biggl({t^2}+  \Bigl(1+ \frac{\alpha^2}{4\ell^2}\Bigr)z^2\biggr)\, \Delta_t -\frac{\alpha^2z}{4\ell}\,(t\partial_t +z\partial_z)^2\, \Delta_t\ . \label{3six4}\eeqa
We again see that  $K^0_\star$ agrees with its commutative analogue $K^0$, while  $K^+_\star$ and  $K^-_\star$  are  deformations of  $K^+$ and $K^-$, (\ref{KlngfG}),  containing infinite order polynomials in $\partial_t$.  As before, the $AdS_2$ Killing vectors are recovered in the commutative limit,  $K^\mu_\star\rightarrow K^\mu$ as $\alpha\rightarrow 0$.

The expressions (\ref{3six4}) for the Killing vectors on   $ncAdS_2$ can be used to examine another  limit of interest,   $z\rightarrow 0$, which  corresponds to the boundary of  $ncAdS_2$.  In that limit $\Delta_t \Phi\rightarrow \frac z{\ell}\partial_t \Phi|_{z=0}$ and $S_t \Phi\rightarrow \Phi|_{z=0}\;$, and so we obtain the commutative result (\ref{kmuFGcrds}),
\be  K_\star^-\rightarrow -\partial_t \quad\ ,\qquad K_\star^0 \rightarrow -t\partial_t\quad\ ,\qquad K_\star^+\rightarrow - t^2\partial_t
\ .\ee
From   $ncAdS_2$ we thus recover the standard form for the global conformal symmetry generators on the boundary.
 We can then say that    $ncAdS_2$ is asymptotically $AdS_2$.  Therefore the $AdS/CFT$ correspondence principal should be applicable.  We explore this possibility in the next section  with the example of  massless scalar field theory.

\section{Massless scalar field theory on  $ncAdS_2$}
\setcounter{equation}{0}

Here we write down an explicit expression for the  field equation for a massless scalar field on  $ncAdS_2$.    Although it describes a free scalar field on   $ncAdS_2$,  the scalar field picks up nontrivial nonlocal interactions after being mapped to the Moyal-Weyl plane.  We  show that these interactions disappear near the boundary. The field equation can be consistently obtained from an action principle upon imposing Dirichlet boundary conditions, and this is because we find no non-commutative  corrections to the boundary term from variations of the action.

Say  $ \Phi^{(0)}$ is now a massless scalar field on  $AdS_2$.  The standard $SO(2,1)$ invariant action can be written in terms of Poisson brackets with the embedding coordinates
\be  S[\Phi^{(0)}]=\frac 1{2\ell_0}\int_{AdS_2} d\mu\,\{X^{\mu},\Phi^{(0)}\}\{X_\mu,\Phi^{(0)}\}\;\;,\label{cmtvactn}\ee
where $ d\mu$ is an invariant integration measure on $AdS_2$.  When written in terms of canonical coordinates it becomes
\beqa  S[\Phi^{(0)}]&= &\frac 1{2\ell_0}\int_{ {\mathbb{R}}^2} dxdy\,\Bigl\{ \Bigl( y \partial_y\Phi^{(0)}+ \partial_x \Phi^{(0)}\Bigr)^2 \;+\;\ell_0^2\,(\partial_y\Phi^{(0)})^2\Bigr\}\ ,\label{clsfacc}
\eeqa
while it reduces to (\ref{clmsfa}) when written in terms of  Fefferman-Graham coordinates.

Upon promoting $\Phi^{(0)}$ to a field $\hat\Phi$   on  $ncAdS_2$, there is an obvious generalization of (\ref{cmtvactn}) to an $SO(2,1)$ invariant action for  $\hat\Phi$.  It is
\be  S_{nc}[\hat \Phi]=-\frac 1{2\ell}{\rm Tr}\,[\hat X^\mu,\hat \Phi][\hat X_\mu,\hat \Phi]\;\;,\label{mslssclrfldactn}\ee
where Tr denotes a trace operation. Here for simplicity we assume that the  $ncAdS_2$ scale parameter is the same as the commutative one, $\ell=\ell_0$; i.e., $\ell$ has no $\alpha^2$ dependence. (\ref{mslssclrfldactn}) can be mapped to an action  on the  Moyal-Weyl plane
\be  S_{nc}[\Phi]=-\frac 1{2\ell\alpha^2} \int_{ {\mathbb{R}}^2} dxdy\,[{\cal X}^\mu,\Phi]_\star\star [{\cal X}_{\mu},\Phi]_\star\;\;,\label{Sclrfldactn}\ee
where the trace has been replaced by  $ \frac 1{\alpha^2} \int_{ {\mathbb{R}}^2} dxdy$.
  Upon applying   (\ref{smnlsofXmu}) and (\ref{sixpt59}) in the appendix one gets
\be  S_{nc}[\Phi]=\frac 1{2\ell\alpha^2}\int_{ {\mathbb{R}}^2} dxdy\,\biggl\{- [y,\Phi]_\star^{\,2}\;+\;[e^x,\Phi]_\star\,\,\Bigl[ \,y\star e^{-x}\star y +k (k+1)\,e^{-x} \, ,\,\Phi\Bigr]_\star  \biggr\} \;,\ee
where we are ignoring all boundary terms because for the moment we shall only be concerned with the field  in the bulk.
(Boundary affects are taken into account below.) Using (\ref{strcmtrsFi})  this becomes
\be
S_{nc}[\Phi]=\frac 1{2\ell}\int_{ {\mathbb{R}}^2} dxdy\,\biggl\{(\partial_x\Phi)^2 + \Delta_y\Phi \,\biggl(y^2 \Delta_y\Phi+2 y\partial_x S_y\Phi-\frac{\alpha^2}4\partial_x^2\Delta_y\Phi\biggr)\;+\;\alpha^2\Bigl(k+\frac 12\Bigr)^2(\Delta_y\Phi)^2\biggr\}\;,
\ee
up to boundary terms.  Upon integrating by parts and using $$\int_{ {\mathbb{R}}^2} dxdy\,\biggl\{\Bigl(\partial_x S_y\Phi\Bigr)^2 - \frac{\alpha^2}4\,\Bigl(\partial_x\Delta_y\Phi\Bigr)^2-(\partial_x\Phi)^2\biggr\}=0\;,$$ it simplifies to
\be
S_{nc}[\Phi]=\frac 1{2\ell}\int_{ {\mathbb{R}}^2} dxdy\,\biggl\{ \Bigl( y \Delta_y\Phi+ \partial_x S_y\Phi\Bigr)^2 \;+\;\Bigl(\frac {\alpha^2}4+{\ell^2}\Bigr)(\Delta_y\Phi)^2\biggr\}\ .\label{LntrmsofDltS}
\ee  This is an an explicit expression for the bulk action in terms of the canonical coordinates.
In terms of  Fefferman-Graham coordinates, the action is
\be  S_{nc}[\Phi]=\frac 1{2}\int_{ {\mathbb{R}}\times  {\mathbb{R}}_+} dt dz\,\frac{1}{z^2}\,\biggl\{\Bigl( \frac{\ell t} z \Delta_t\Phi- (t\partial_t+z\partial_z) S_t\Phi\Bigr)^2 \;+\;\Bigl(\frac {\alpha^2}4+{\ell^2}\Bigr)(\Delta_t\Phi)^2\biggr\} \label{actnnfgcrs}\ .\ee
(\ref{clsfacc}) and  (\ref{clmsfa}) are  recovered from the commutative limit,  $\alpha\rightarrow 0$, of (\ref{LntrmsofDltS}) and (\ref{actnnfgcrs}), respectively.

We note that as one approaches the boundary $z=0$, the action density goes  to that of a massless scalar field on commutative  $AdS_2$, with a rescaled time parameter $t$.  Using $\Delta_t \Phi\rightarrow \frac z{\ell}\partial_t \Phi|_{z=0}$ and $S_t \Phi\rightarrow \Phi|_{z=0}\;$ as  $z\rightarrow 0$, the integrand  in (\ref{actnnfgcrs}) goes to\footnote{In passing from canonical coordinates to Fefferman-Graham coordinates we used the commutative formulas (\ref{mp2pncra}) (with the natural change $l_0 \rightarrow l$).  On the other hand, one can re-absorb the factor in (\ref{lgrdennrbnd}) by re-scaling $t$ (or, $z$) in a quantum (or noncomutative) version of  (\ref{mp2pncra}). The commutative limit, of course, of this transformation must coincide with  (\ref{mp2pncra}). Because this does not seem to bring any radical simplification, we will keep on using the commutative change of variables (\ref{mp2pncra}).}
  \be \Bigl(1+\frac {\alpha^2}{4\ell^2}\Bigr)\, (\partial_t \Phi)^2\;+\;(\partial_z \Phi) ^2  \label{lgrdennrbnd}\;,\ee
as compared to the integrand in (\ref{clmsfa}).
This means that the commutative free field equation is recovered near the boundary, again with a rescaled coordinate,
  \be\Bigl(1+\frac {\alpha^2}{4\ell^2}\Bigr) \partial^2_t \,\Phi  \,\;+\;  \partial_z^2 \Phi\;\rightarrow\; 0\;\;,\qquad{\rm as}\;\; z\rightarrow 0\;,\label{zgszrofldeq}\ee
and so  $\Phi$ satisfies the equation for a massless scalar field on an asymptotically $AdS_2$ space.

The field equation for $\Phi$ can be written down for all $z$.
Variations in $\Phi$ in (\ref{Sclrfldactn}) yield
\beqa \delta S_{nc}[\Phi]&=&-\frac 1{2\ell\alpha^2} \int_{ {\mathbb{R}}^2} dxdy\,\Bigl([{\cal X}^\mu,\delta\Phi]_\star\star [{\cal X}_\mu,\Phi]_\star +[{\cal X}^\mu,\Phi]_\star\star [{\cal X}_\mu,\delta\Phi]_\star\Bigr )\cr &&\cr
&=&-\frac 1{2\ell\alpha^2} \int_{ {\mathbb{R}}^2} dxdy\,\Bigl(2[{\cal X}^\mu,\delta\Phi]_\star\star [{\cal X}_\mu,\Phi]_\star +[[{\cal X}^\mu,\Phi]_\star \,, [{\cal X}_\mu,\delta\Phi]_\star ]_\star\Bigr)
\cr &&\cr
&=&\frac 1{\ell\alpha^2} \int_{ {\mathbb{R}}^2} dxdy\,\delta\Phi\star[{\cal X}^\mu, [{\cal X}_\mu,\Phi]_\star]_\star\cr &&\cr &&\qquad\quad -\;\frac 1{2\ell\alpha^2} \int_{ {\mathbb{R}}^2} dxdy\,\Bigl(2[{\cal X}^\mu,\delta\Phi\star [ {\cal X}_\mu,\Phi]_\star ]_\star +[[{\cal X}^\mu,\Phi]_\star \,, [{\cal X}_\mu,\delta\Phi]_\star ]_\star\Bigr)\ .\cr&&\label{vrtnactnmwp}\eeqa
From the first term,
the field equation in the bulk is
 \be  [{\cal X}^\mu,[{\cal X}_{\mu},\Phi]_\star]_\star=0\ .\label{XXfI}\ee

 The  remaining two terms [last line in (\ref{vrtnactnmwp})] are only defined on the boundary.  This is since the Moyal star commutator of any two functions ${\cal F}$ and ${\cal G}$ on the Moyal-Weyl plane is a total divergence.  Following (\ref{bndryfsp}) in the appendix we can write the integral of $ [{\cal F},{\cal G}]_\star$ over $D$ as $\int_{\partial D}\,( {\cal V}_xdx +{\cal V}_ydy)$,
where  $\partial D$ is the    boundary of $D$.  $ {\cal V}_x $ and $ {\cal V}_y $ are computed up to order $\alpha^2$ in (\ref{vxandvy}). For us the boundary is located at $z=0$, and so $\int_{\partial D}\,( {\cal V}_xdx +{\cal V}_ydy)= \int\, {\cal V}_t |_{z=0}\; dt$, where  $ {\cal V}_t=\frac{\ell}{z} {\cal V}_y$.
To compute ${\cal V}_t$ for the first boundary term in (\ref{vrtnactnmwp})
 we  set ${\cal F}$ and  ${\cal G}$ in  (\ref{bndryfsp}) equal to ${\cal X}^\mu$ and $\delta\Phi\star [{\cal X}_{\mu},\Phi]_\star $, respectively, and then sum over $\mu$.
  At leading order in $\alpha$, ${\cal V}_t=-\alpha^2\ell\,\delta\Phi\partial_z\Phi$.  This is the commutative result.  After some work we get that the $\alpha^2$  corrections to this result  go like $z^n$, $n\ge 1$, which then vanish after setting $z=0$.
To compute ${\cal V}_t$ for the second boundary term in (\ref{vrtnactnmwp})
 we  set ${\cal F}$ and  ${\cal G}$ in  (\ref{bndryfsp}) equal to  $[{\cal X}^{\mu},\Phi]_\star $ and $[{\cal X}_{\mu},\delta\Phi]_\star $, respectively, and then sum over $\mu$.   We find that all contributions to ${\cal V}_t$ go like $z^n$, $n\ge 1$, which once again vanish after setting $z=0$.   We thus get that all non-commutative corrections to the boundary terms vanish.  Although  we have only checked this to order $\alpha^2$ we expect that the result is true to all orders since they involve higher order derivatives which will produce higher powers in $z$ in  ${\cal V}_t$.  The boundary term in (\ref{vrtnactnmwp}) is then just the commutative answer
 \be - \int dt \,(\partial_z\Phi\,\delta\Phi)\Big|_{z=0}\ .\label{btnvrtnS} \ee
 This means that we can fix the boundary value of the field
 \be \phi_0(t)=\Phi(0,t) \;,\label{bndrevlufi}\ee and  the variational problem is well defined for Dirichlet boundary conditions.

Alternatively, the field equation in the bulk can be found directly from the Lagrangian density (\ref{LntrmsofDltS})
 with the help of the identities
\beqa \int_{ {\mathbb{R}}^2} dxdy\,\Bigl(\Delta_y A(x,y)B(x,y)+ A(x,y)\Delta_y B(x,y)\Bigr)&=&0\ ,\cr &&\cr
  \int_{ {\mathbb{R}}^2} dxdy\,\Bigl(S_y A(x,y)B(x,y)- A(x,y)S_y B(x,y)\Bigr)&=&0 \ ,\eeqa
which are valid up to boundary terms. Note that the first identity shows that under integration, $\Delta_y$ behaves as the usual derivative satisfying the Leibnitz rule.
  Then the field equation  following from  (\ref{LntrmsofDltS}) is
  \be (  \Delta_yy+ \partial_x S_y )\,( y \Delta_y +\partial_x S_y)\,\Phi  \;+\;\Bigl(\frac {\alpha^2}4+{\ell^2}\Bigr)\, \Delta_y^2\Phi=0\;,\ee
  or in  Fefferman-Graham coordinates,
   \be \Bigl(  \ell\,\Delta_t \frac{t}z- (t\partial_t+z\partial_z) S_t \Bigr)\,\Bigl(\ell\,\frac{t}z \Delta_t- (t\partial_t+z\partial_z) S_t\Bigr)\,\Phi  \;+\;\Bigl(\frac {\alpha^2}4+{\ell^2}\Bigr)\, \Delta_t^2\Phi=0\ .\label{3sixseven}\ee
 In both limits  $\alpha\rightarrow 0\,$ and  $z\rightarrow 0\,$, (\ref{3sixseven}) reduces to a second order differential equation. In the former, we  recover the commutative answer (\ref{cmtvsffe}), while
in the latter, (\ref{3sixseven}) reduces to the previously obtained  result near the boundary (\ref{zgszrofldeq}).
 Although  (\ref{3sixseven}) contains infinitely many orders in derivatives with respect to $t$, it is only   second order in derivatives in $z$ (just as in the commutative case).  Then it can be solved given sufficient data at the $AdS$ boundary, which we do to leading order in $\alpha^2$ in the next section.

\section{Leading order solutions and the $CFT_1$ correspondence}
\setcounter{equation}{0}

Here we compute the on-shell action and resulting two-point function for the boundary theory to leading order in the noncommutativity parameter.
Expanding the field equation (\ref{3sixseven}) up to the
leading order correction in $\alpha^2$ gives
\be \Box\,\Phi-\frac{\alpha^2}{12\ell^2}\,\Bigl\{t \partial_t+z^2\partial_t^2+9z\partial_z+2zt\partial_z\partial_t+3z^2\partial_z^2 \Bigr\} \partial_t^2\Phi+{\cal O}(\alpha^4)=0\ .\label{slnalfaexp}\ee
Using standard techniques,\cite{DHoker:2002nbb} one can write down a solution to (\ref{slnalfaexp}) in terms of the boundary value of the field (\ref{bndrevlufi}), which we can define  to be independent of $\alpha^2$.  We denote the solution by $\Phi_{sol}[\phi_0]$.
We expand $\Phi_{sol}[\phi_0]$ in powers of $\alpha^2$ about the commutative solution $\Phi^{(0)}$, satisfying (\ref{cmtvsffe}),
\be\Phi_{sol}[\phi_0]=  \Phi^{(0)}+\alpha^2 \Phi^{(1)}+\cdots +\alpha^{2M} \Phi^{(M)}+\cdots \;\ee
  $\Phi^{(0)}$ is solved in (\ref{frstrdrntsln})  using the boundary-to-bulk propagator.
From (\ref{slnalfaexp}), the leading order non-commutative correction  $\Phi^{(1)}$  satisfies
\be \Box\,\Phi^{(1)}=\frac{1}{12\ell^2}\,\Bigl\{t \partial_t+z^2\partial_t^2+9z\partial_z+2zt\partial_z\partial_t+3z^2\partial_z^2 \Bigr\} \partial_t^2\Phi^{(0)}\ .\label{lwstrdrsfe}\ee
After using (\ref{frstrdrntsln}) on the right hand side  we get
$$ \Box\,\Phi^{(1)}= \frac z{2 \pi\ell^2} \int dt'{\cal F}(t,t',z)\phi_0 (t')\;,$$
\be {\cal F}(t,t',z)= \frac{ z^6-   ( t +35 t') (t-t')\,z^4-5 ( t-17 t') (t-t')^3 \,z^2-3 (t+3 t') (t-t')^5}{
   \Bigl((t-t')^2+z^2\Bigr)^5}\ .\ee
   We now apply the bulk-to-bulk propagator\cite{Fronsdal:1974ew},\cite{Burgess:1984ti},\cite{Inami:1985wu} 
     \be  G(z,t;z',t')=\frac 1{2\pi}\tanh^{-1}\Bigl(\frac{2 z z'}{z^2+z'^2+(t- t')^2}\Bigr)\ ,\ee 
 satisfying   $ \;\,\Box \,G(z,t;z',t')=-\delta(z-z')\delta(t-t')\;, $  
    to obtain an  integral expression  for $\Phi^{(1)}$
   \be \Phi^{(1)}(z,t)= -\frac 1{2 \pi\ell^2}\int_0^\infty dz'{z'}\int dt' \int dt''\, G(z,t;,z',t'){\cal F}(t',t'',z')\phi_0 (t'')\;,\label{ldngordrctn}
    \ee
  This procedure can in principal be repeated to get any higher order correction $ \Phi^{(M)}$ to the commutative field.

We next use (\ref{frstrdrntsln}) and (\ref{ldngordrctn}) to  compute the on-shell action.  For this purpose it is convenient to re-express the action (\ref{Sclrfldactn}) as
\be  S_{nc}[\Phi]=\frac 1{2\ell\alpha^2}\int dxdy\,\Phi\star [{\cal X}^\mu,[{\cal X}_{\mu},\Phi]_\star]_\star -\frac 1{2\ell\alpha^2} \int dxdy\,[{\cal X}^\mu,\Phi\star [{\cal X}_{\mu},\Phi]_\star]_\star\ .\label{altntvactn}\ee
From  (\ref{XXfI}), the first term vanishes on-shell.  The  remaining term  is only defined on the boundary since the Moyal star commutator is a total divergence. We can once again use (\ref{bndryfsp}) in the appendix to compute it  up to order $\alpha^2$ in (\ref{vxandvy}).  Setting ${\cal F}$ and  ${\cal G}$ in  (\ref{bndryfsp}) equal to ${\cal X}^\mu$ and $\Phi\star [{\cal X}_{\mu},\Phi]_\star $, respectively, and summing over $\mu$, we get $ {\cal V}_t=\frac{\ell}{z} {\cal V}_y=\alpha^2\ell\,\Phi\partial_z\Phi$ at leading order in $\alpha$.
  After some work we get that the $\alpha^2$  corrections to this result  go like $z^n$, $n\ge 2$, which then vanish after setting $z=0$.
This means that the expression for the on-shell action receives no non-commutative corrections (at least, at  order $\alpha^2$)
\beqa  S_{nc}[\Phi_{sol}[\phi_0]]&=&-\frac 1{2\ell\alpha^2} \int dxdy\,[{\cal X}^\mu,\Phi\star [{\cal X}_\mu,\Phi]_\star]_\star\Big|_{\Phi=\Phi_{sol}[\phi_0]}\cr &&\cr
&=&-\frac 12\int dt \;\Phi_{sol}[\phi_0] \; \partial_z\Phi_{sol}[\phi_0] \Big|_{z=0}\ .\label{Sncphisol} \eeqa
This is identical to the commutative result (\ref{cmtvBndTurm}).

It remains to substitute the solution (\ref{frstrdrntsln})  and (\ref{ldngordrctn}) into the action (\ref{Sncphisol}).  This gives
\beqa && S_{nc}[\Phi_{sol}[\phi_0]]\;=\;-\frac 12\int dt \int dt'\,\phi_0(t) \,\biggl( \partial_z K(z,t;t') \Big|_{z=0}\phi_0 (t')\cr &&\cr &&\qquad\qquad -\;\frac {\alpha^2 }{2 \pi\ell^2}\int_0^\infty dz'{z'} \int^\infty_{-\infty} dt''\, K(z',t;t'){\cal F}(t',t'',z')\phi_0 (t'')+{\cal O}(\alpha^4)\biggr)
\cr &&\cr &&\cr &=&-\frac{1}{2\pi}\int dt \int dt'\,\phi_0(t) \,\biggl(\frac{ \phi_0(t')}{(t-t')^2}-\frac {\alpha^2 }{2 \pi\ell^2}\int_0^\infty dz'\int^\infty_{-\infty} dt''\,  \frac{{{z'^2}\, {\cal F}(t',t'',z')\,\phi_0 (t'')}}{{z'}^2+(t- t')^2}+{\cal O}(\alpha^4)\biggr)
\cr &&\cr &&\cr &=&-\frac{1}{2\pi}\int dt \int dt'\,\phi_0(t)\phi_0(t')\biggl(\frac{1 }{(t-t')^2}-\frac {\alpha^2 }{2 \pi\ell^2}\int_0^\infty dz' \int^\infty_{-\infty} dt''\,  \frac{{{z'^2}\,{\cal F}(t'',t',z')}}{{z'}^2+(t- t'')^2}+{\cal O}(\alpha^4)\biggr)\;,\cr&&\label{ncrsn2nsa}\eeqa
where we used the identity
$ \partial_z G(z,t;z',t') \Big|_{z=0}= K(z',t;t')$.   The the second term in parenthesis in (\ref{ncrsn2nsa}) is the leading non-commutative correction. It
can be exactly computed using the integral
\be\int_0^\infty dz \int^\infty_{-\infty} dt'' \,\frac{{{z^2}\,{\cal F}(t'',t',z)}}{{z}^2+(t- t'')^2}=\frac{\pi /4 }{(t-t')^2}\ .\ee
This result means that the on-shell action merely undergoes an overall rescaling
\beqa  S_{nc}[\Phi_{sol}[\phi_0]]&=&-\frac{1}{2\pi}\int dt \int dt'\,\phi_0(t)\phi_0(t')\biggl(\Bigr(1-\frac {\alpha^2 }{8 \ell^2}\Bigr)\frac{1 }{(t-t')^2}+{\cal O}(\alpha^4)\biggr)\ .\cr&&\eeqa
Then from the $AdS/CFT$ correspondence (\ref{adscft}),   $n-$point correlation functions of quantum mechanical operators ${\cal O}(t)$ on the one-dimensional boundary
 also undergo an overall rescaling at leading order in the noncommutativity parameter.  For
 the two-point function we get
\be <{\cal O}(t){\cal O}(t ')>\;=\;-\frac{1}{2\pi}\Bigl(1-\frac {\alpha^2 }{8 \ell^2}\Bigr)\frac{1 }{(t-t')^2}+{\cal O}(\alpha^4)\ .\label{fnlrslt2ptfn}\ee
  Recall that at the beginning of section six, we fixed  $\ell$ equal to the commutative length scale $\ell_0$.  If $\ell$ instead depends on  $\alpha$, we should   replace $\ell$ in the leading order correction in (\ref{fnlrslt2ptfn}) by $\ell_0$.

\section{Concluding Remarks}

  We have shown that  $ncAdS_2$ has a commutative boundary, implying that $ ncAdS_2$ is assymptotically $AdS_2$.  Then from general arguments the $AdS/CFT$ correspondence should be applicable. We explicitly demonstrated this by computing the two-point function on the boundary associated with the massless scalar field on   $ncAdS_2$.  The dynamics for the scalar field contains nontrivial nonlocal interactions, which is evident from  the Moyal-Weyl plane description.  These interactions vanish at the $ncAdS_2$ boundary.    Our leading order results  show that the introduction of noncommutativity
on the $AdS_2$ space does not affect the boundary conformal theory, other then to  generate a  rescaling of the correlation functions.  The conformal dimension, which is one for the commutative theory is unaffected at leading order in $\alpha^2$.   Higher order computations are feasable. If  the conformal dimension remains one to all orders, the commutative  and non-commutative theory are equivalent within the context of the $AdS/CFT$ correspondence principal.   Our results utilized the isometry preserving commutation relations (\ref{adstoocrs}) which defines   $ncAdS_2$.  Different results may follow from other deformations of anti-de Sitter space.  This was found recently for a $\kappa$-deformed $AdS_2$ space-time.\cite{Gupta:2017xex}  There the conformal dimension was a  nontrivial function of the noncommutativity parameter.

Concerning the issue of disconnected time-like boundaries of $AdS_2$,\cite{Strominger:1998yg} we find that Euclidean  $ncAdS_2$ selects a single boundary.  This is because the boundary in this system is described in terms of  states of a particular discrete series representation $D^+(k)$ (or  $D^-(k)$), which have a lowest (or highest) state.  As a result, the eigenvalues of the radial coordinate operator $\hat r$  has a lower (or upper) bound, namely zero, while the boundary corresponds to the $\hat X^1$ eigenvalue going to $+\infty$ (or  $-\infty$).

A number of generalizations of our work are possible.   Among them is the addition  of a mass term Tr $\hat \Phi^2$, or interaction terms  Tr $\hat \Phi^M$ to the action  (\ref{mslssclrfldactn}) of the scalar field on $ncAdS_2$. This will introduce further nonlocal interactions in the Moyal-Weyl plane description,  and is  likely to lead to non-commutative corrections to the Breitenlohner-Freedman bound.\cite{Breitenlohner:1982bm} The examination of other fields on  $ncAdS_2$, such as spinors, gauge fields and spin-two fields is another very natural extension of our work.  A Dirac operator has been proposed for  $ncAdS_2$,\cite{Fakhri:2011zz} which can be utilized in writing down an action for spinors.
 Gauge fields on $AdS_2$ were recently examined in \cite{Mezei:2017kmw}
 and it may be possible to check whether or not they have non-commutative extensions. Within the context of the non-commutative theory, the spin-two fields should represent quantum gravity fluctuations.
 The massless scalar field examined in this article required no Gibbons-Hawking-York boundary term, nor holographic renormalization, as fields were asymptotically finite.  Such simplifications most likely will not apply  for the other field theories on  $ncAdS_2$. 
 
  Generalizations to  $ncAdS_{d+1},\,d>1$ should prove even more challenging. In this case there is no prefered choice for the Poisson brackets and their corresponding quantization, both of which will necessarily break the $AdS_{d+1}$ isometry group, and hence the conformal symmetry on the boundary.  For example, it may be desirable to posit the Poisson bracket  (\ref{rtcnclycngt}), since it states that the time is canonically conjugate to the $CFT_d$ energy scale.  However for  $d>1$ this Poisson bracket breaks the full Lorentz (or Euclidean) symmetry on the boundary. In another example Poisson brackets on $AdS_4$ were given in \cite{Chaney} (section 5.4.2)  which  broke  the $SO(3, 2)$ isometry group to $SO(3) \otimes SO(2)$.   Thus  more complicated result for the correlation functions are   expected for $d>1$.

\bigskip
\bigskip

\appendix
\appendice{\Large{\bf $\quad$ Some properties of the Moyal-Weyl star product}}

 Given two functions ${\cal F}$ and $ {\cal G}$ on the  Moyal-Weyl plane spanned by $(x,y)$,
 their
 star product is defined by
 \be  [ {\cal F}\star  {\cal G} ](x,y) =  {\cal F}(x,y)\,\exp{\Bigl\{\,\frac{i\alpha}2 \,(\overleftarrow{ { \partial_ x }}\,
\overrightarrow{ {\partial_ y }}\,-\,\overleftarrow{ { \partial_ y }}\,
\overrightarrow{ {\partial_ x }}
)\,
\Bigr\}}\; {\cal G}(x,y)\ . \label{dffstrprd}\ee
This definition leads to the identities
the following identities for the Moyal-Weyl star product
\beqa
 {\cal F}(x)\star={\cal F}\Bigl(x+\frac {i\alpha}2\overrightarrow{ {\partial_ y }}\Bigr)\ , &&\star  {\cal F}(x)= {\cal F}\Bigl(x-\frac {i\alpha}2\overleftarrow{ {\partial_ y }}\Bigr)\ ,\cr &&\cr
  {\cal G}(y)\star={\cal G}\Bigl(y-\frac {i\alpha}2\overrightarrow{ {\partial_ x }}\Bigr)\ , &&\star {\cal G}(y)={\cal G}\Bigl(y+\frac {i\alpha}2\overleftarrow{ {\partial_ x }}\Bigr)\ .\label{MWidntes}
\eeqa

A property of the integral of the Moyal-Weyl star product of two functions  ${\cal F}$ and $ {\cal G}$ on the  Moyal-Weyl plane is
\be\int_{ {\mathbb{R}}^2} dxdy \,{\cal F}\star  {\cal G}   =\int_{ {\mathbb{R}}^2} dxdy\,{\cal F} {\cal G}\;+\;{\rm boundary}\;{\rm terms}\ .\label{sixpt59}\ee
Correspondingly,
the Moyal star commutator is a total divergence.  The integral of a star commutator  of any two functions ${\cal F}$ and ${\cal G}$ on the Moyal-Weyl plane  can then be written as a boundary integral\be  \int_{D} dxdy\, [{\cal F},{\cal G}]_\star(x,y)=\int_{D} dxdy\,[\partial_x{\cal V}_y-\partial_y{\cal V}_x](x,y)=\int_{\partial D}\,( {\cal V}_xdx +{\cal V}_ydy)\;,\label{bndryfsp}\ee
where $D $ is some two-dimensional domain, with   boundary $\partial D$.  Up to order $\alpha^2$, $ {\cal V}_x $ and $ {\cal V}_y $ are
\beqa   {\cal V}_x&=&i\alpha \biggl(-\partial_x{\cal F}\,{\cal G} +\frac{\alpha^2}{24}\Bigl(\partial_x^3{\cal F\,}\partial^2_{y}{\cal G}+\partial_x\partial^2_{y}{\cal F}\,\partial^2_{x}{\cal G}-2\partial_x^2\partial_{y}{\cal F}\,\partial_{x}\partial_{y}{\cal G}\Bigr)+{\cal O}(\alpha^4)\biggr)\ , \cr &&\cr {\cal V}_y&=&i\alpha \biggl(-\partial_y{\cal F}\,{\cal G} +\frac{\alpha^2}{24}\Bigl(\partial_y^3{\cal F}\,\partial^2_{x}{\cal G}+\partial^2_x\partial_{y}{\cal F}\,\partial^2_{y}{\cal G}-2\partial_x\partial^2_{y}{\cal F}\,\partial_{x}\partial_{y}{\cal G}\Bigr)+{\cal O}(\alpha^4)\biggr)\ .\cr&& \label{vxandvy}\eeqa

\bigskip
\bigskip
{\Large {\bf Acknowledgments} }

\noindent
 We are very grateful to A.P. Balachandran, A. Chaney, M. Kaminski, C.  Uhlemann and J. Wu for valuable discussions.

 \end{document}